# Environment-assisted quantum control of a solid-state spin via coherent dark states

Jack Hansom[1*], Carsten H. H. Schulte[1*], Claire Le Gall[1], Clemens Matthiesen[1], Edmund Clarke[2], Maxime Hugues[3], Jacob M. Taylor[4] & Mete Atatüre[1]

[1] *Cavendish Laboratory, University of Cambridge, JJ Thomson Avenue, Cambridge CB3 0HE, UK*

[2] *EPSRC National Centre for III-V Technologies, University of Sheffield, Sheffield, S1 3JD, UK*

[3] *CNRS-CRHEA, rue Bernard Grégory, 06560 Valbonne, France*

[4] *Joint Quantum Institute/National Institute of Standards and Technology, Gaithersburg, MD 20899, USA*

\* These authors contributed equally to this work.

**Understanding the interplay between a quantum system and its environment lies at the heart of quantum science and its applications. To-date most efforts have focused on circumventing decoherence induced by the environment by either protecting the system from the associated noise[1-5] or by manipulating the environment directly[6-9]. Recently, parallel efforts using the environment as a resource have emerged, which could enable dissipation-driven quantum computation[10,11] and coupling of distant quantum bits[12,13]. Here, we realize the optical control of a semiconductor quantum-dot spin by relying on its interaction with an adiabatically evolving spin environment. The emergence of hyperfine-induced, quasi-static optical selection rules enables the optical generation of coherent spin dark states without an external magnetic field. We show that the phase and amplitude of**

**the lasers implement multi-axis manipulation of the basis spanned by the dark and bright states, enabling control via projection into a spin-superposition state. Our approach can be extended, within the scope of quantum control and feedback[14,15], to other systems interacting with an adiabatically evolving environment.**

Techniques for controlling spins often rely upon a well-defined Zeeman splitting due to a static external magnetic field. This field also controls the selection rules of the optical transitions enabling, for example, optical single-shot spin readout[16-18] and fast spin manipulation[2,19] in self-assembled quantum dots (QDs). Unfortunately, these two capabilities require different alignments of the external field with respect to the growth direction of the QD: a field along the growth direction of a QD (Faraday configuration) provides cycling transitions for spin readout, while a magnetic field applied perpendicular to the growth direction (Voigt configuration) enables optically-driven spin control. In an effort to achieve both readout and control on a QD spin qubit, a large, rapidly switchable, multi-axis magnetic field would be the ideal toolbox. While the quantisation axis due to a static field in Voigt configuration can in principle be converted to that in Faraday configuration via the optical Stark effect, the prohibitively large laser power required to achieve this renders the scheme impractical. We instead consider the smallest effective field we can devise: that due to the local fluctuating nuclear spins within the QD. They give rise to an effective magnetic field for an electron spin in the QD, known as the Overhauser (OH) field (Fig. 1a), with a dispersion of 10-30 mT and a many-microsecond correlation time[20-22]. The hyperfine interaction therefore lifts the Kramers degeneracy and provides a quantisation axis for the electron spin, which in turn leads to temporally quasi-stable optical selection rules due to the quasi-static nature of the nuclei (Fig. 1b). This OH field is small enough to allow for Stark tilting the quantisation axis for optical read-out feasibly using laser pulses, provided coherent control of the spin can be achieved using only the OH field for quantisation.

Working with a single, negatively-charged indium arsenide QD (Methods) at low external magnetic field, a single optical line is observed (Fig. S 3) as the linewidth exceeds the OH-induced Zeeman splitting. However, this single line is due to four different transitions (Fig. 1b), which lead to non-trivial correlations between the outgoing photons in time. Specifically, a low-power intensity autocorrelation measurement $g^{(2)}(\tau)$ detuned from the bare resonance (Fig. 1c) shows, in addition to the well-known antibunching of photons near zero time delay, the presence of bunching, i.e. $g^{(2)}(\tau) > g^{(2)}(\infty)$. Further, the degree of bunching increases as the laser is detuned by one linewidth, indicating that one spin state is preferentially addressed. This reveals the spin dynamics[23] characteristic of fast optical spin pumping via spin-flip Raman processes, showing the nuclear environment provides the spin-selective access needed for optical spin control, even though spectrally resolving the transitions is not possible. A model of the four level system in Fig. 1 agrees well with the sub-linewidth detuning dependence shown in the inset of Fig. 1c and yields an OH field dispersion of $18 \pm 1$ mT, corresponding to a spin splitting of 0.42 times the measured absorption linewidth ($\Gamma_{abs}=2.23\Gamma$) (SI section 1).

We now seek to prepare the spin state coherently via optical pumping. A conceptually simple extension is the generation of spin coherences via two-colour coherent population trapping (CPT)[24]. Specifically, the destructive interference of two different excitation paths driven via lasers at two different optical frequencies selects an energy-superposition state of the spin, which will not be excited ('dark' state); optical decay of the other spin superposition (the 'bright' state) eventually shelves the spin into the dark state. Thus, in a conventional three-level Λ-system, the formation of a coherent spin dark state manifests itself as a decrease in absorption at the two photon resonance (TPR). The magnitude and spectral width of this absorption dip are directly related to the laser intensity and the ground state coherence time[24-26]. In our slowly varying system both the TPR and the selection rules are determined by the evolving OH field magnitude and orientation, respectively. Fortunately, a reasonable fraction of the OH field distribution gives rise to a level structure which admits coherent superpositions of energy eigenstates (coherent dark states) (SI section 6), as schematically depicted in Fig. 2a. In order to detect the spectral signature of CPT, we record the fluorescence intensity while scanning the frequencies of two linearly and orthogonally polarised resonant lasers (shown as red (H) and green (V) arrows in Fig. 2a), each set to $\Omega=0.224\ \Gamma$. We spectrally filter and record the red-detuned phonon-assisted[27] QD fluorescence (Methods), allowing arbitrary flexibility in the excitation laser frequencies and polarisations, paramount to laser background-free detection of CPT with near-degenerate ground states. Figure 2b shows the theoretically predicted QD absorption as a function of $\Delta_{1,2}$, where $\Delta_i$ is the detuning of laser *i* from resonance (SI section 6). Resonant excitation of QDs yields the well-studied dynamic nuclear spin polarisation effect[6,7,21] which is modelled here empirically as an added contribution to the splitting of the ground states. A simulation with 400-MHz (0.83 $\Gamma_{abs}/2\pi$) ground state splitting is in good qualitative agreement with the experimental data at zero field, shown in Fig. 2c. Specifically, the reduced absorption due to CPT is visible for equal laser frequencies ($\Delta_1=\Delta_2$) in the experimental data. An analytical calculation of the ground state decay rates shows that this feature is due to the optical dressing of the electron spin states leading to the existence of a coherent dark state in the form of

$$|\Psi\rangle \propto \alpha|\uparrow,\hat{n}(t)\rangle - e^{i\varphi}\beta|\downarrow,\hat{n}(t)\rangle, \tag{1}$$

where the unit vector $\hat{n}(t)$ is defined by the orientation of the adiabatically evolving OH field, ↑ and ↓ are spin projections defined along $\hat{n}(t)$, $\alpha$ and $\beta$ are effective Rabi frequencies and $\varphi$ depends on the relative phase of the two laser fields. The projection of this dark state in the hyperfine-dictated, bare basis is given by the ratio of absolute Rabi frequencies (SI section 6). In parallel, the state in equation (1), together with an orthogonal bright state, spans the dressed basis.

The low visibility of the CPT feature is a result of its intermittent nature: only a fraction of the OH field distribution provides the desired Λ-scheme for CPT. Therefore, stable coherent dark states are generated intermittently and sustained for a finite period of time, dictated by the slow dynamics of the nuclei. This leads to an apparent reduction of the CPT visibility in our time-averaged experiment and is in contrast to other systems displaying a reduction of the CPT visibility caused by fast spin dephasing timescales on the order of the radiative lifetime of the optical transitions[28,29]. The observation of a pronounced decrease in absorption here verifies the adiabatic evolution of the OH field with respect to the dressed state dynamics, resulting in relatively long-lived coherent dark states.

Panels d and e of Fig. 2 show line-cuts along $\Delta_1+\Delta_2=0$, for two values of applied field in the Faraday configuration. Applying 18.4 mT external magnetic field (Fig. 2e), comparable in magnitude to the OH field, leads to the breakdown of the hyperfine-assisted Λ-system and therefore to the disappearance of the CPT dip. The magnetic field dependence of the CPT visibility is summarized in Fig. 2f, corrected for saturation and spin pumping effects (SI section 4). Additionally, the dependence of this visibility on the electron spin coherence time is probed straightforwardly by operating the QD device in the cotunneling regime (SI section 2), i.e. where the interaction of the QD spin with the Fermi sea of the back contact[30] leads to reduced spin coherence and the the CPT feature disappears.

The coherent dark state is determined by the complex Rabi frequencies of the optical transitions in the Λ-system and hence, a relative phase change of the lasers imposes a rotation of the dressed basis about $\hat{n}(t)$[31]. For adiabatic changes, the electron follows the dark state and is hence rotated in the bare basis. In order to expose the optical phase dependence of the coherent dark states, we impose a non-adiabatic phase jump after preparing a dark state and measure its effect on the time-resolved QD fluorescence. This rapid phase manipulation leads to a new rotated dressed basis in which the original dark state gains a finite bright state component, thus generating fluorescence. Subsequent photon scattering projects the electron spin into the new dark state[31]. Figures 3a and 3b depict the experimental setup and the QD level structure, respectively, used to evidence the phase-dependent fluorescence by imprinting a phase jump on one of the excitation lasers using an electro-optic modulator (EOM) (Methods).

A representative time-resolved measurement for a sudden phase jump is shown in Fig. 3c, where the bottom panel shows the photon detection events, while the upper panels display the applied EOM voltage and its derivative. The phase jump leads to a sharp dip, synchronised with the falling edge of the electrical pulse. This dip is caused by the detuning of the laser during the phase change, $\Delta_{eff} = \partial \phi / \partial t$, as depicted in the middle panel, proportional to a change in instantaneous frequency. Following the sharp dip, we see a transient in fluorescence as the electron gains a bright state component after a phase-dependent rotation of the dressed basis (cf Bloch spheres). This transient arises only from an abrupt phase-jump-induced change in fluorescence, and is therefore conditional on the QD spin being in a coherent dark state (left-most Bloch spheres). This enables us to observe the phase dependence of the dark state with intermittent CPT and the characteristic dark-state formation timescale.

The demonstration of the phase-dependence of the dark state, along with the relative laser amplitude control, yields multi-axis control of the dressed basis. In a quantum control picture[14,15], the dressed basis acts as a laser-defined measurement and preparation basis which allows for control of the electron spin by projection. The state information in this picture is provided directly by the fluorescence proportional to the bright state component prior to repumping into the new dark state after a few optical cycles. To evidence such quantum control, we vary the amplitude of the phase jump described earlier and record the subsequent photon emission. The plot in Fig. 4 displays the intermittent fluorescence intensity as a function of phase-jump amplitude controlled by the EOM voltage and corrected for background signal (SI section 5). Since the intermittent fluorescence intensity is proportional to the bright state component the spin state gains after the phase jump (cf. Bloch spheres in Fig. 4), the measured sinusoidal dependence evidences controlled rotation of the dressed basis. After subsequent optical pumping into the rotated dark state, this protocol implements quantum control of the electron spin about the quasi-static quantisation axis, $\hat{n}(t)$, enabled by the hyperfine interaction with the nuclei.

We note that the dressed basis is defined with respect to the unknown nuclear environment. Nonetheless, we can keep the spin in a known dark state (Eq. 1) and can manipulate it deterministically by controlling the laser parameters. Remarkably, this lack of knowledge does not translate to dephasing in the time-averaged protocol. These results imply a straightforward extension to time-synchronised protocols, in which the heralded preparation of a dark state could be performed and utilised within the correlation time of the OH field. The modest magnitude of the OH field used here to construct a heralded Voigt configuration will allow for rapid switching into the orthogonal geometry through the application of a pulsed spin-selective AC Stark shift. Together, this combines the dual requirements of coherent spin control and single-shot spin readout with a single QD, which remains elusive to-date. Environment-assisted CPT naturally lends itself to protocols which demand the initialisation of the spin into a coherent superposition state and sub-linewidth ground-state Zeeman splitting, such as photonic cluster state generation[32]. In such a protocol, the electron spin is prepared in a superposition state and precesses at a rate slower than the radiative lifetime, while a string of subsequently scattered photons form the computational resource. Realising this protocol requires two additional capabilities: First, the creation of a coherent spin state needs to be heralded via single-shot optical readout[17,18]. Second, the disentangling of the electron spin from the generated photons needs to be implemented via a projective measurement of the spin in the laboratory frame. Both requirements can be achieved with the phonon sideband detection technique demonstrated here.

**Methods**

**Sample:** We use a sample grown by molecular beam epitaxy containing a single layer of self-assembled InAs/GaAs QDs in a GaAs matrix, embedded in a Schottky diode for charge state control. The Schottky diode structure comprises an n+-doped layer 35 nm below the QDs and a 5-6 nm thick partially transparent titanium layer evaporated on top of the sample surface. This device structure allows for deterministic charging of the QDs and shifting of the QD exciton energy levels via the DC Stark effect. 20 pairs of GaAs/AlGaAs layers forming a distributed Bragg reflector extend to 205 nm below the QD layer for increased collection efficiency in the spectral region between 960 nm and 980 nm. Spatial resolution and collection efficiency are enhanced by a zirconia solid immersion lens in Weierstrass geometry positioned on the top surface of the device.

**Phonon-sideband filtering:** Typically for QDs at 4.2 K, the phonon sideband accounts for around 10% of the resonance fluorescence spectrum and has a bandwidth of ~2 nm. Reflected laser and QD resonance fluorescence is incident on a pulse-shaping grating in a configuration close to a Littman configuration. The dispersed beam is focussed and a sharp edged mirror placed in the Fourier plane selectively reflects part of the spectrum. This reflection is sent back onto the grating to re-collimate the beam and coupled into a single-mode fibre. The saturated QD photon detection rate through the filtering stage is 13 kHz on a fibre-coupled avalanche photo diode, which corresponds to ~2.2% of the typical count rate obtained with a cross polarisation technique. Accounting for the polarisation dependence and efficiency of the grating, the fibre coupling efficiency and losses due to the optics, this corresponds to collecting ~7.1% of the emission spectrum of the QD. For powers lower than $P_{sat}$, the laser reflection is undetectable and the signal-to-background ratio is typically larger than 100 and limited by detector dark counts.

**Electro-optic phase modulation:** The total phase change in Fig. 3c corresponds to the difference in phase before and after the jump (electrical fall time < 2 ns). At B = 0 T the selection rules evolve according to the OH field, therefore the lasers do not address a single transition selectively and a relative phase of the lasers is not mapped directly to a relative phase of the complex Rabi frequencies. We apply a small magnetic field (8.4 mT) to split the excited states, while the phase-modulated excitation laser is set to higher power than the fixed-phase laser ($P_1 = 4 \cdot P_2 = P_{sat}/5$ where $P_{sat} = \Gamma^2/2$ time-averaged), leading to dark states with a *z* component in the bare basis (SI section 7). An 80-MHz detuning between the lasers compensates for the ground state splitting induced by the magnetic field, as illustrated in Fig. 3b. Rectangular pulses of variable amplitude drive the EOM, the same electrical signal is correlated with the time-resolved resonance fluorescence from the QD. The height of the transient is measured ~2 ns after the phase jump, in order to avoid phase-change-dependent effects unrelated to CPT. Imperfections in the electrical pulses lead to intermittent fluorescence of smaller amplitude that is not related to the transformation of a dark state into a bright state. This background signal decays on a much larger timescale and can be measured in an experiment with only one laser (SI section 5). Subtracting the background from the two-laser measurement yields the increase in fluorescence due to projection onto the bright state (Fig. 4).

**Acknowledgements**

We gratefully acknowledge financial support by the University of Cambridge, the European Research Council ERC Consolidator Grant agreement no. 617985, EU-FP7 Marie Curie Initial Training Network S$^3$NANO, the NSF-funded Physics Frontier Center at the Joint Quantum Institute, and ARO MURI award no. W911NF0910406. The authors also acknowledge J.C. Barnes, G. Solomon, M.J. Stanley, R.H.J. Stockill, and E. Waks for fruitful discussions and technical assistance. J.M.T. thanks the Atomic, Mesoscopic and Optical Physics Group at the Cavendish Laboratory for their fine hospitality during his stays.

**Author contributions**

J.H., C.H.H.S., J.M.T. and M.A. devised the experiments. J.H., C.H.H.S and C.L.G. performed the experiments and analysed the data. J.H., C.H.H.S., C.L.G., C.M., J.M.T. and M.A contributed to the discussion of the results and the manuscript preparation. J.M.T. performed the theoretical modelling shown in Fig. 2. J.H. performed theoretical modelling of the data shown in Fig. 1. E.C. and M.H. grew the sample. C.M. processed the devices.

Correspondence and requests for materials should be addressed to J.M.T. (jmtaylor@umd.edu) and M.A. (ma424@cam.ac.uk).

**Competing financial interests**

The authors declare no competing financial interests.

30. Dreiser, J., Atatüre, M., Galland, C., Müller, T., Badolato, A. & Imamoglu, A. Optical investigations of quantum dot spin dynamics as a function of external electric and magnetic fields, *Phys. Rev. B* **77**, 075317 (2008).

31. Yale, C. G., *et al.* All optical control of a solid state spin using coherent dark states, *PNAS* **110**, 7595 (2013).

32. Lindner, N.H. & Rudolph, T. Proposal for pulsed on-demand sources of photonic cluster state strings, *Phys. Rev. Lett.* **103**, 113602 (2009).

**FIG. 1. Optical spin access via environment-dictated quantisation axis. (a)** The hyperfine interaction between the resident electron spin (black arrow) and a large bath of nuclear spins (orange arrows) is modelled as a classical effective magnetic field, the Overhauser field (OH, large orange arrow). **(b)** Level structure and selection rules for different orientations of the OH field. The curved arrows represent dipole-allowed transitions, the straight arrow represents the excitation laser used in (c). **(c)** Low power ($\Omega$=0.224 $\Gamma$) resonance fluorescence intensity autocorrelation (blue line) fitted with an exponential decay (red line). The data is post-selected for a rms-detuning of ~0.23 $\Gamma$ (SI section 1). The inset shows the bunching amplitude for different post-selected detunings. The red curve is a spin pumping simulation with an OH field dispersion of 18±1 mT. Horizontal axes are in units of radiative transition linewidth $\Gamma$ = 216±2 MHz and lifetime $\tau$ = 737±6 ps. Error bars represent the standard error in the amplitudes of fits to exponential decays for each measurement.

**FIG. 2. Environment-assisted coherent population trapping. (a)** Two lasers, each set to $\Omega$=0.224 $\Gamma$, with orthogonal linear polarization (H, V) and variable detuning, are used to excite the QD. **(b)** Simulation of two-laser absorption of the QD at zero magnetic field and without spectral wandering. Experimentally measured values of laser powers, radiative lifetime and OH field dispersion are used. A hyperfine-induced ground state splitting of 400 MHz gives the best fit to the data. **(c)** Two-laser measurement of QD absorption with experimental parameters corresponding to (b). **(d,e)** Linecuts across $\Delta_1+\Delta_2$=0 (dashed black line shown in panel (c)) for two different applied magnetic fields in the Faraday configuration. The corresponding linecut extracted from the simulation of (b) with spectral wandering is shown in red. **(f)** Magnetic field dependence of the visibility of the spectral signature of CPT, after correcting for saturation and incoherent spin pumping effects (SI). The measured OH field dispersion is shown as a vertical

orange line. The error bars are calculated by propagation of both the min-max values of the count rates around $\Delta_1=\Delta_2=0$, and standard errors in the amplitude of Voigt fits to single-laser line shapes used for normalisation.

**FIG. 3. Phase-dependence of coherent dark states. (a)** Experimental setup used to measure the effect of fast phase jumps on QD absorption. A single laser is split into two paths, one is phase-modulated with an electro-optic modulator (EOM) and the other is frequency shifted to compensate for the ground state Zeeman splitting under 8.4 mT (applied to lift the excited state degeneracy for this experiment). Photon detection events are recorded and correlated with voltage pulses sent to the EOM; AOM: Acousto-Optic Modulator; PBS: Polarising Beam Splitter; APD: Avalanche Photodiode; λ/2: Half-wave plate. **(b)** QD level structure at an arbitrary OH field and 8.4 mT external field, with phase and frequency differences corresponding to the experimental setup described in (a). **(c)** Example time-resolved fluorescence measurement (lower graph) when an electrical pulse (upper graph) is sent to the EOM. The middle graph shows the phase change rate, corresponding to an effective frequency shift. The Bloch spheres depict qualitatively the effect of the phase jump on the electron spin state at different times in the cycle, both in the hyperfine-dictated and dressed bases (*B* is bright state, *D* is dark state).

**FIG. 4. Quantum control of an electron spin via state projection**. Amplitude of the intermittent fluorescence extracted from fits to exponential decays, as a function of voltage pulse amplitude applied to the EOM, normalised to the mean count rate. The blue curve is a sinusoidal fit, and the error bars represent the standard deviation of the background-subtracted (SI) counts in each measurement. $V_\pi$ is the EOM voltage required for a π phase shift. The Bloch spheres

depict the effect of the corresponding phase jump on an electron spin prepared in the dark state. The transparent vector represents the initially prepared dark state, and the opaque vector shows the final state.

**Figure 1**

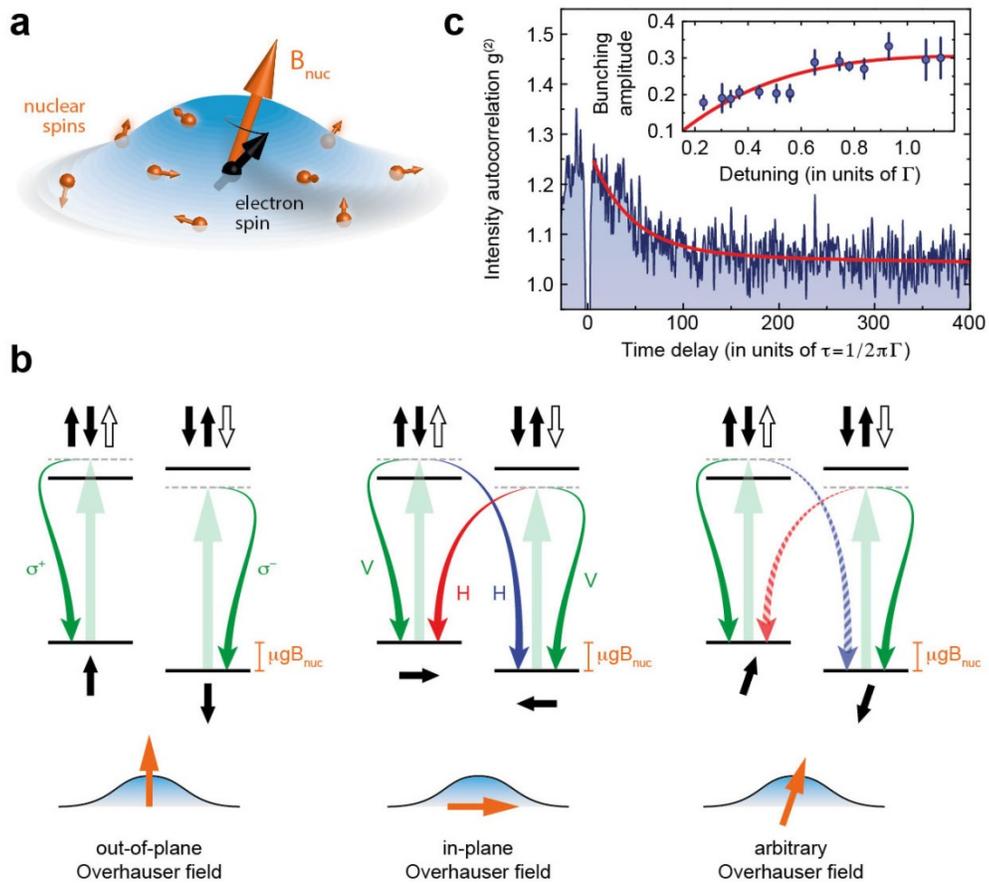

**Figure 2**

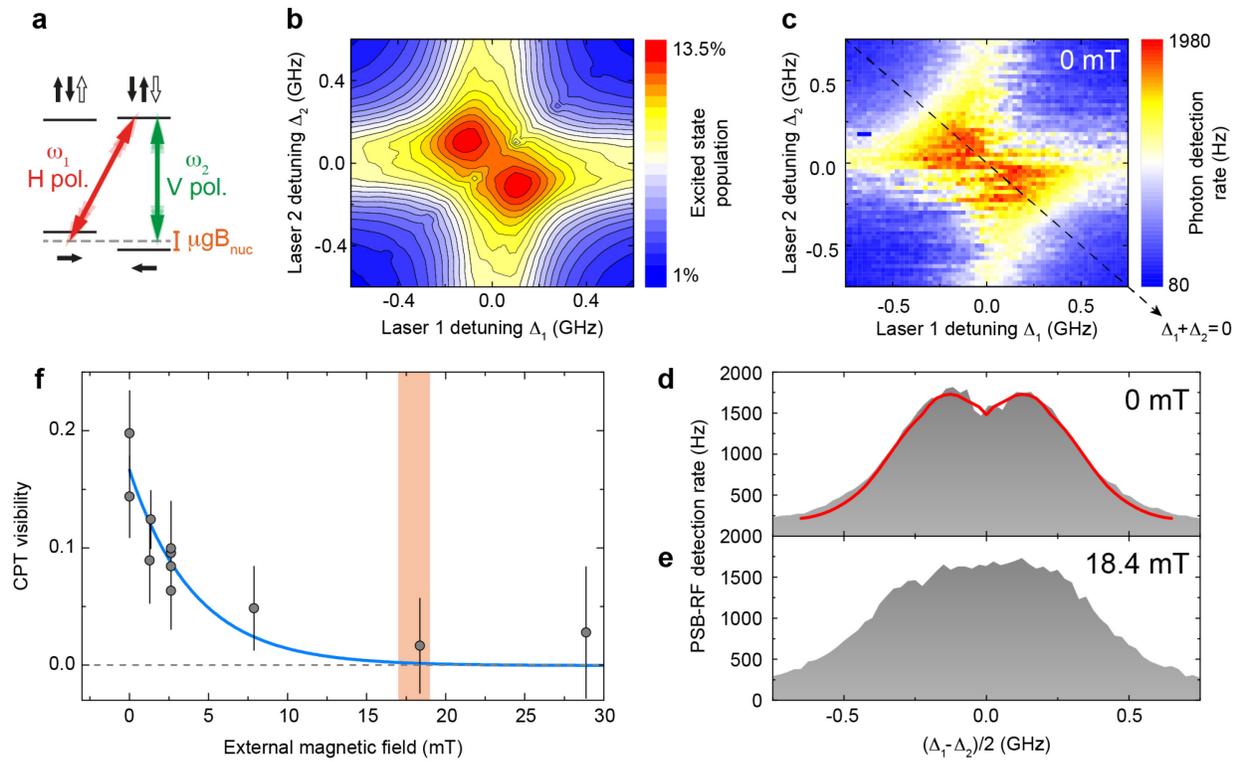

**Figure 3**

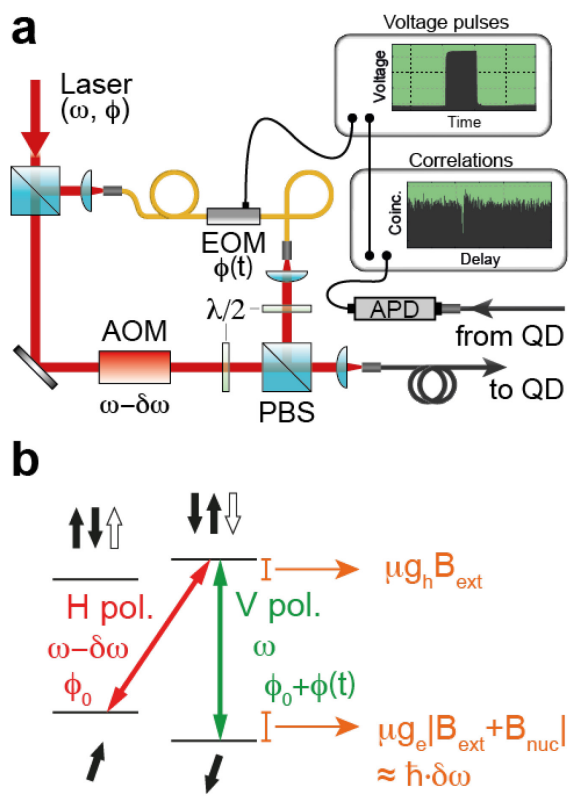
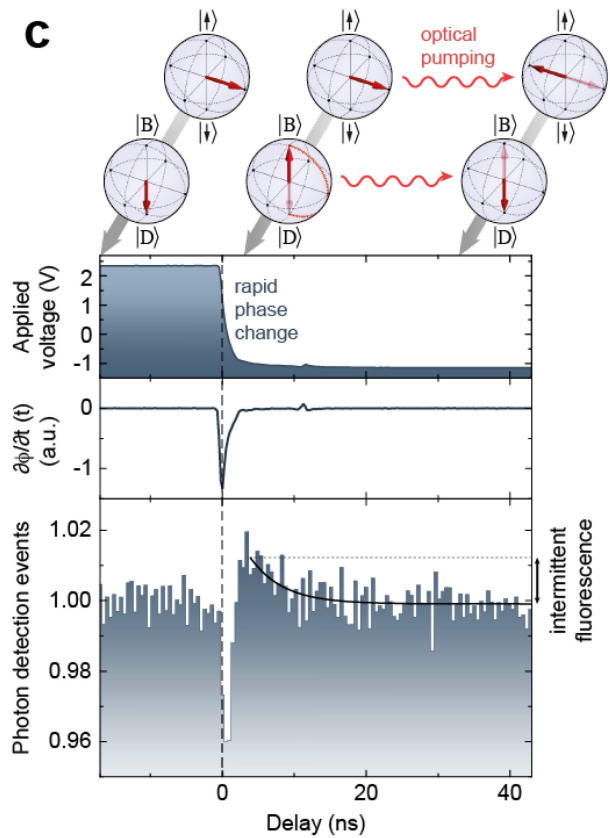

**Figure 4**

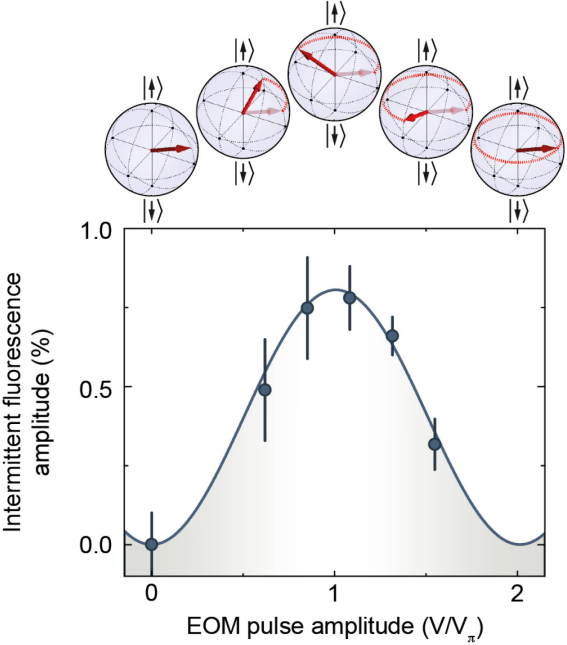